\documentclass[graybox]{svmult}

\usepackage{type1cm}                           
\usepackage{makeidx}                           
\usepackage{graphicx}                          
\usepackage{multicol}                          
\usepackage[stable,bottom]{footmisc}           
\usepackage{amsmath}
\usepackage{newtxtext}
\usepackage{newtxmath}                         
\usepackage{gensymb}
\usepackage{graphicx}
\usepackage{tikz}

\makeindex             

\begin{document}
\title*{A non-reflective boundary condition for LBM \\ based on the assumption of non-equilibrium symmetry}
\titlerunning{A non-reflective boundary condition for LBM}
\author{R. Euser and C. Vuik}
\institute{
	R. Euser \at Delft Institute of Applied Mathematics, \email{ralf.euser@gmail.com} \and
	C. Vuik  \at Delft Institute of Applied Mathematics, \email{c.vuik@tudelft.nl}}
%
%
\maketitle
\abstract*{In this study a new type of non-reflective boundary condition (NRBC) for the Lattice Boltzmann Method (LBM) is proposed;
the Non-equilibrium Symmetry Boundary Condition (NSBC). The idea behind this boundary condition is to utilize
the characteristics of the non-equilibrium distribution function to assign values to the incoming populations.
A simple gradient based extrapolation technique and a far-field criterion are used to predict the macroscopic fluid variables.
To demonstrate the non-reflective behaviour of the NSBC, two different tests have been carried out,
examining the capability of the boundary to absorb acoustic waves respectively vortices.
The results for both tests show that the amount of reflection generated by the NSBC is nearly zero.
}
\abstract{In this study a new type of non-reflective boundary condition (NRBC) for the Lattice Boltzmann Method (LBM) is proposed;
the Non-equilibrium Symmetry Boundary Condition (NSBC). The idea behind this boundary condition is to utilize
the characteristics of the non-equilibrium distribution function to assign values to the incoming populations.
A simple gradient based extrapolation technique and a far-field criterion are used to predict the macroscopic fluid variables.
To demonstrate the non-reflective behaviour of the NSBC, two different tests have been carried out,
examining the capability of the boundary to absorb acoustic waves respectively vortices.
The results for both tests show that the amount of reflection generated by the NSBC is nearly zero.
}
\section{Introduction} \label{sec:int}
In many fluid dynamics applications the region of interest comprises of only a small subdomain in space and time.
When modelling such applications using numerical methods, ideally, one would like to isolate this region,
as to minimize computational expenses and to allow for a sufficiently fine grained discretization.
Isolating this region often requires advanced boundary treatment, in which continuity of the flow field is assumed.
In other words, the amount of energy being reflected at the boundary has to be zero.
This is where the so-called \emph{Non-Reflective Boundary Conditions} (NRBC's) come into practise.
Focusing on compressible flow solvers like LBM, the NRBC's can be divided into two different groups.
The first group, known as the \emph{Characteristic Boundary Conditions} (CBC's), aims at canceling out reflections by suppressing any incoming waves.
The second group, the \emph{Absorbing Layer} (AL) approach, uses a layer of several nodes thick to absorb any outgoing waves.
Over the years, various efforts have been made to model such NRBC's with LBM.
In 2006 Chikatamarla et al. proposed Grad's approximation for missing data \cite{grad2006},
in which incoming populations are assigned values based on a low-dimensional sub-manifold in distribution function space.
Shortly thereafter Kam et al. \cite{kam2006, kam2007} published about the use of NRBC's for aeroacoustics simulations,
in which he compares several boundary treatments based on extrapolation \cite{chen1996, li2002},
filtering \cite{giatonde1999, visbal2001} and absorbing layers \cite{taasan1995, freund1997}.
By the end of 2008 Izquierdo and Fueyo \cite{izquierdo2008} proposed an LBM formulation for the Characteristic Boundary Condition (CBC),
based on the one-dimensional (LODI) characteristics of the Euler equations and their extension to the Navier-Stokes equivalent (NS-CBC) \cite{nordstrom1995, poinsot1992}.
Following the work of Hu \cite{hu1996, hu2001, hu2005, hu2008}, Najafi-Yazdi and Mongeau \cite{yazdi2009, yazdi2012} developed
a direction independent AL-NRBC based on the Perfectly Matched Layer (PML) approach.
In 2013 Schlaffer \cite{schlaffer2013} presented an extensive research on NRBC's, in which he introduced the Impedence Boundary Condition (IBC).
To continue with the CBC developments, Heubes et al. \cite{heubes2014} proposed a linear combination between Thompson's boundary conditions \cite{thompson1987} and the LODI relations.
Comparing the approaches of Izquierdo and Heubes, Puig-Ar\`anega et al. \cite{puig2015} found that the LODI equations become inappropriate when the dimensionality of the flow increases.
Consequently, Jung et al. \cite{jung2015} developed a two-dimensional generalization of the CBC,
by recovering the transverse and viscous terms in the characteristics analysis \cite{yoo2005, yoo2007, lodato2008}.
Extending on the above approaches, Wissocq et al. \cite{wissocq2017} were able to improve the numerical stability of the CBC
at high Reynolds numbers by taking advantage of a regularized collision scheme \cite{latt2006}.
In this study a new type of boundary concept is proposed to approximate non-reflective flow behavior at the boundary.
The idea of this concept is to utilize the characteristics of the non-equilibrium distribution function to assign values to the incoming populations.
A simple gradient extrapolation technique coupled with a far-field reference vector is used to predict the macroscopic fluid variables.
%
%
%
\section{The Boltzmann Transport Equation} \label{sec:bte}
Based on kinetic theory, the Boltzmann Transport Equation (BTE) \eqref{eq:bte} decribes the statistical behavior of molecular motion
inside a system by using a seven-dimensional \emph{probability density function} $f$,
also referred to a \emph{particle distribution function} (PDF) or simply \emph{distribution function} when using the concept of \emph{fictituous particles}:
\begin{equation} \label{eq:bte}
	\frac{\partial{f}}{\partial{t}}+\xi_\alpha
	\frac{\partial{f}}{\partial{x_\alpha}}+\frac{F_\alpha}{\rho}
	\frac{\partial{f}}{\partial{\xi_\alpha}}=\Omega(f)
\end{equation}
where $f$ is a function of time $(t)$ space $(\vec{x})$ and \emph{velocity space} $(\vec{\xi})$.
Whenever a medium relaxes towards steady state, the solution of the BTE becomes the \emph{equilibrium distribution function} $f^{\mathrm{eq}}$ (EDF):
\begin{equation} \label{eq:feq_0}
	f^{\mathrm{eq}}(\rho,\vec{u},\theta,\vec{\xi}) =
	\frac{\rho}{(2 \pi \theta)^{d/2}}
	e^{-\frac{{\vert\vec{\xi}-\vec{u}\vert}^2}{2 \theta}}
\end{equation}
where all quantities are non-dimensional; $\rho$ is the density and $\theta$ is the temperature, equal to $R T / u_0$,
in which $R$ and $u_0$ are respectively the gas constant and characteristic velocity; $\vec{u}$ is the macroscopic velocity of the medium and $d$ the number of spatial dimensions.
A key component of equation \eqref{eq:bte} is the \emph{collision operator} $\Omega(f)$,
which represents all possible ways in which particles can collide with one another:
\begin{equation} \label{eq:bgk}
	\Omega(f) = -\frac{1}{\tau}(f - f^{\mathrm{eq}})
\end{equation}
where $\tau$ is the \emph{relaxation time}, a direct function of the transport coefficients of a medium, such as viscosity and heat diffusivity.
The \emph{macrosocopic moments} like \emph{mass density} \eqref{eq:cf_0} and \emph{momentum density} \eqref{eq:cf_1}
can be obtained by integrating the moments of $f$ respectively $f^{\mathrm{eq}}$ over the $d$-dimensional velocity space:
\begin{alignat}{2}
	\rho=&\int{f\,d^d\xi}&&=\int{f^{\mathrm{eq}}\,d^d\xi} \label{eq:cf_0} \\
	\rho u_\alpha=&\int{\xi_\alpha f\,d^d\xi}&&=\int{\xi_\alpha f^{\mathrm{eq}}\,d^d\xi} \label{eq:cf_1}
\end{alignat}
\section{The Lattice Boltzmann Method} \label{sec:lbm}
\begin{figure}[ht]
	\begin{minipage}{0.25\textwidth}
		\centering
		\begin{tikzpicture}[scale=2]
			\draw (-0.5, -0.5) rectangle (0.5, 0.5);
			\draw[->, >=stealth, line width=1] (0.0, 0.0) -- (0.5, 0.0);		
			\draw[->, >=stealth, line width=1] (0.0, 0.0) -- (-0.5, 0.0);		
			\draw[->, >=stealth, line width=1] (0.0, 0.0) -- (0.0, 0.5);		
			\draw[->, >=stealth, line width=1] (0.0, 0.0) -- (0.0, -0.5);		
			\draw[->, >=stealth, line width=1] (0.0, 0.0) -- (0.5, 0.5);		
			\draw[->, >=stealth, line width=1] (0.0, 0.0) -- (-0.5, -0.5);		
			\draw[->, >=stealth, line width=1] (0.0, 0.0) -- (-0.5, 0.5);		
			\draw[->, >=stealth, line width=1] (0.0, 0.0) -- (0.5, -0.5);		
			\draw (0.0, 0.0) node[fill=white, shape=circle, inner sep=0pt] {$\vec{c}_0$};
			\draw (0.6, 0.0) node[fill=white, shape=circle, inner sep=0pt] {$\vec{c}_1$};
			\draw (-0.6, 0.0) node[fill=white, shape=circle, inner sep=0pt] {$\vec{c}_2$};
			\draw (0.0, 0.6) node[fill=white, shape=circle, inner sep=0pt] {$\vec{c}_3$};
			\draw (0.0, -0.6) node[fill=white, shape=circle, inner sep=0pt] {$\vec{c}_4$};
			\draw (0.57, 0.57) node[fill=white, shape=circle, inner sep=0pt] {$\vec{c}_5$};
			\draw (-0.57, -0.57) node[fill=white, shape=circle, inner sep=0pt] {$\vec{c}_6$};
			\draw (-0.57, 0.57) node[fill=white, shape=circle, inner sep=0pt] {$\vec{c}_7$};
			\draw (0.57, -0.57) node[fill=white, shape=circle, inner sep=0pt] {$\vec{c}_8$};
		\end{tikzpicture}
	\end{minipage}
	\hspace{0.5cm}
	\begin{minipage}{0.25\textwidth}
		\begin{center}
			\begin{tabular}{ c c c }
				$i$ & $\vec{c}_i$ & $w_i$ \\
				\hline
				0 & $(0,0)$ & $4/9$ \\
				1--2 & $(\pm1,0)$ & $1/9$ \\
				3--4 & $(0,\pm1)$ & $1/9$ \\
				5--8 & $(\pm1,\pm1)$ & $1/36$ \\
			\end{tabular}
		\end{center}
	\end{minipage}
	\hspace{0.5cm}
	\begin{minipage}{0.25\textwidth}
		\centering
		\begin{tikzpicture}[scale=0.8]
			\def\var{1.5}
			\def\scale{0.3}
			\shade[inner color=gray, outer color=white, opacity=1.0] (0.0, 0.0) rectangle (3.0, 3.0);
			\draw[color=gray, step=1.0] (0.0, 0.0) grid (3.0, 3.0);
			\draw (\var,\var) node[fill=white, circle, scale=\scale]{};
			\draw (\var+1,\var) node[fill=black, circle, scale=\scale]{};
			\draw (\var-1,\var) node[fill=black, circle, scale=\scale]{};
			\draw (\var,\var+1) node[fill=black, circle, scale=\scale]{};
			\draw (\var,\var-1) node[fill=black, circle, scale=\scale]{};
			\draw (\var+1,\var+1) node[fill=black, circle, scale=\scale]{};
			\draw (\var-1,\var-1) node[fill=black, circle, scale=\scale]{};
			\draw (\var-1,\var+1) node[fill=black, circle, scale=\scale]{};
			\draw (\var+1,\var-1) node[fill=black, circle, scale=\scale]{};
			\draw[color=white, ->, >=stealth] (\var+0.4,\var) -- (\var+0.1, \var);
			\draw[color=white, ->, >=stealth] (\var-0.4,\var) -- (\var-0.1, \var);
			\draw[color=white, ->, >=stealth] (\var,\var+0.4) -- (\var, \var+0.1);
			\draw[color=white, ->, >=stealth] (\var,\var-0.4) -- (\var, \var-0.1);
			\draw[color=white, ->, >=stealth] (\var+0.4,\var+0.4) -- (\var+0.1, \var+0.1);
			\draw[color=white, ->, >=stealth] (\var-0.4,\var-0.4) -- (\var-0.1, \var-0.1);
			\draw[color=white, ->, >=stealth] (\var-0.4,\var+0.4) -- (\var-0.1, \var+0.1);
			\draw[color=white, ->, >=stealth] (\var+0.4,\var-0.4) -- (\var+0.1, \var-0.1);
			\draw[color=black, ->, >=stealth] (\var+0.6,\var) -- (\var+0.9, \var);
			\draw[color=black, ->, >=stealth] (\var-0.6,\var) -- (\var-0.9, \var);
			\draw[color=black, ->, >=stealth] (\var,\var+0.6) -- (\var, \var+0.9);
			\draw[color=black, ->, >=stealth] (\var,\var-0.6) -- (\var, \var-0.9);
			\draw[color=black, ->, >=stealth] (\var+0.6,\var+0.6) -- (\var+0.9, \var+0.9);
			\draw[color=black, ->, >=stealth] (\var-0.6,\var-0.6) -- (\var-0.9, \var-0.9);
			\draw[color=black, ->, >=stealth] (\var-0.6,\var+0.6) -- (\var-0.9, \var+0.9);
			\draw[color=black, ->, >=stealth] (\var+0.6,\var-0.6) -- (\var+0.9, \var-0.9);
		\end{tikzpicture}
	\end{minipage}
	\caption{D2Q9 model -- Lattice configuration (left) and exchange between lattices (right)}
    \label{fig:d2q9}
\end{figure}
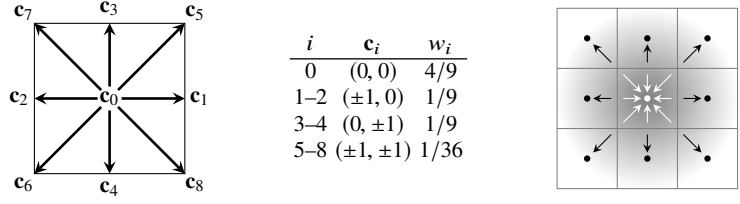

Based on the Lattice Boltzmann equation (LBE) \eqref{eq:lbe_0}, the Lattice Boltzmann method (LBM)
is a direct discretization of the BTE in both time, space and velocity space:
\begin{equation} \label{eq:lbe_0}
	f_i(\vec{x}+\vec{c}_i\Delta{t},t+\Delta{t})-f_i(\vec{x},t) = \Omega_i(\vec{x},t)
\end{equation}
The method constructs numerical approximations by iteratively \emph{streaming} and \emph{colliding} discrete distribution functions $f_i$,
confined by the discrete velocities $\vec{c}_i$ of a \emph{lattice} (Figure~\ref{fig:d2q9}).
By introducing compact notation for \eqref{eq:lbe_0} and substituting the collision operator \eqref{eq:bgk}, the LBE can be rewritten as:
\begin{equation} \label{eq:lbe_1}
	f_i^*=f_i-\frac{\Delta{t}}{\tau}\left(f_i - f_i^{\mathrm{eq}}\right)
\end{equation}
where $f_i^*$ are the discrete \emph{post-collision} distribution functions and $\Delta{t}$ is the discrete time step.
The discrete equilibrium distribution function $f_i^{\mathrm{eq}}$ is given by:
\begin{equation} \label{eq:feq_1}
	f_i^{\mathrm{eq}} = w_i\rho\left(1+\frac{\vec{u}\cdot\vec{c}_i}{c_s^2}+
	\frac{(\vec{u}\cdot\vec{c}_i)^2}{2c_s^4}-\frac{\vec{u}\cdot\vec{u}}{2c_s^2}\right)
\end{equation}
where $c_s=\frac{1}{\sqrt{3}}$ is the LBM speed of sound and $w_i$ are the discrete \emph{weights},
associated with the lattice velocities $\vec{c}_i$.
\section{The Multiple-Relaxation-Time collision model} \label{sec:mrt}
To increase the accuracy and stability of the current solution, the so-called Multiple-relaxation-time (MRT) collision model has been added.
By relaxing the \emph{velocity moments} of $\vec{f}$ at different rates, rather than relaxing $\vec{f}$ itself at a single rate,
the MRT collision model is capable of modelling a large range of Reynolds numbers. Similar to \eqref{eq:lbe_1} the MRT LBE is given by:
\begin{equation} \label{eq:mrt_lbe}
	\vec{f}^*=\vec{f}-\vec{M}^{-1}\vec{S}\left(\vec{m} - \vec{m}^{\mathrm{eq}}\right)\Delta{t}
\end{equation}
where $\vec{m}$ and $\vec{m}^{\mathrm{eq}}$ are respectively the \emph{velocity moments} and the \emph{equilibrium velocity moments}:
\begin{equation} \label{eq:mrt_vm}
	\vec{m}=\vec{M}\cdot\vec{f} \qquad \vec{m}^{\mathrm{eq}}=\vec{M}\cdot\vec{f}^{\mathrm{eq}}
\end{equation}
The quantity $\vec{M}$ is a $Q \times Q$ \emph{transformation matrix}, whose entries can be found by constraining the moments of $\vec{m}$.
Following the Gram-Schmidt (GS) procedure \cite{lbm2017}, $\vec{M}$ can be formed by constructing a set of \emph{mutually orthogonal vectors},
each corresponding to a certain moment of $\vec{f}$. The quantity $\vec{S}$ in \eqref{eq:mrt_lbe} represents the \emph{relaxation matrix}
and is used to relax the different velocity moments. In the case of the GS approach, this matrix has the following diagonal form:
\begin{equation} \label{eq:mrt_rm}
	\vec{S}=\text{diag}\left(C_{\rho},\omega_e,\omega_{\epsilon},C_{j_x},\omega_q,C_{j_y},\omega_q,\omega_{\nu},\omega_{\nu}\right)
\end{equation}
where $\omega_e$ and $\omega_{\epsilon}$ are the enery relaxation rates; $\omega_q$ is the relaxation rate for the energy flux and $\omega_{\nu}$ is the viscous relaxation rate.
The constants $C_{\rho}$, $C_{j_x}$ and $C_{j_y}$ represent the conserved quantities and can be assigned any value.
\section{The non-equilibrium symmetry boundary} \label{sec:nes}
Based upon the approximately symmetrical shape of the non-equilibrium distribution function $f^{\mathrm{neq}}$,
a new type of non-reflective boundary condition (NRBC) has been constructed, known as the Non-equilibrium Symmetry Boundary Condition (NSBC).
The key behind the NSBC is the approximation that the discrete non-equilibrium populations $f_i^{\mathrm{neq}}$ are assumed to be equal
to their anti-symmetric counterparts $f_{\bar{i}}^{\mathrm{neq}}$:
\begin{equation} \label{eq:nes_0}
	f_i^{\mathrm{neq}} = f_{\bar{i}}^{\mathrm{neq}}
\end{equation}
As a result the incoming populations $f_{\mathrm{in}}$ at the boundary nodes 
can be calculated using the non-equilibrium contributions of the outgoing populations $f_{\mathrm{out}}$:
\begin{equation} \label{eq:nes_1}
	f_{\mathrm{in}} = f_{\mathrm{in}}^{\mathrm{eq}} + f_{\mathrm{out}}^{\mathrm{neq}}
\end{equation}
As this approach requires the equilibrium populations $f_{\mathrm{in}}^{\mathrm{eq}}$ to be computed first, correct values for
the macroscopic fluid vector $\vec{m}=(\rho,\vec{u})$ need to be predicted in advance.
Although there exist various approaches to accomplish this \cite{heubes2014, jung2015}, good results were obtained by simply taking
the gradient of $\vec{m}$ along the normal $\vec{n}$ of the boundary,
multiplied by the coefficient $\gamma$, a relaxation parameter used to minimize the amount of reflection.
As for the D2Q9 model $\gamma=0.6$ was found to give the best results:
\begin{equation} \label{eq:nes_2}
	\vec{m}_p =\vec{m} - \gamma\,\partial_{\vec{n}}{\vec{m}}
\end{equation}
To allow for the predicted fluid vector $\vec{m}_p$ to convergence towards a certain reference fluid vector,
a so-called \emph{far-field flow criterion} is introduced, yielding:
\begin{equation} \label{eq:nes_3}
	\vec{m}_c =\left(1-\beta\right)\vec{m}_p + \beta\vec{m}_0
\end{equation}
where the coefficient $\beta$ is the far-field factor and $\vec{m}_0$ the reference fluid vector.
After all boundary populations have been assigned they are corrected by \emph{rescaling} them with respect to $\rho_c$
and \emph{shifting} them with respect to $\vec{u}_c$, as to guarantee conservation of the macroscopic moments:
\begin{equation}
	\tilde{f_i}=f_i-w_i \left[\Delta{\rho}+\vec{c}_i\Delta{\left(\rho \vec{u}\right)} \right] \qquad \text{ with: }
	\begin{cases}
		\Delta{\rho}&=\sum_i{f_i} - \rho_c \\
		\Delta{\left(\rho \vec{u}\right)}&=\sum_i{\vec{c}_i f_i} - \rho_c \vec{u}_c
	\end{cases} \label{eq:nes_4}
\end{equation}
After the correction has been performed, the standard collision procedure can be carried out,
in which there is no distinction between the boundary and the internal fluid.
\section{Test case 1 -- Propagation of acoustic waves} \label{sub:tc_aw}
\begin{figure}[ht]
	\includegraphics[scale=0.85, page=1]{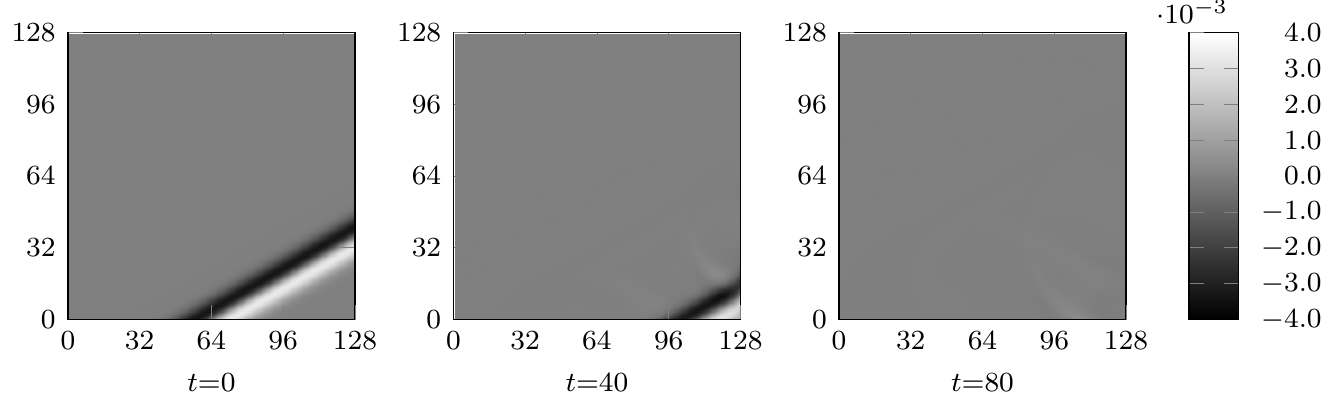}
	\caption{Test case 1: Propagation of a plane wave under an angle -- Pressure results}
	\label{fig:iw_rs}
\end{figure}

Acoustic waves, also known as \emph{sound waves}, are characterised by local pressure variations propagating at a certain speed $c_s$ through a medium.
When considering this medium to be a fluid with negligible viscosity, the propagation of such waves is governed by the \emph{ideal wave equation}:
\begin{equation} \label{eq:we_0}
	\nabla^2 s = \frac{1}{c_s^2}\partial_t^2 s
\end{equation}
A possible solution of \eqref{eq:we_0} is the one-dimensional Gaussian \emph{plane wave} given by:
\begin{align}
	p(x,t) &= \rho_0 c_s^2 \left[1 + s(x,t)\right] \label{eq:pw_p} \\
	u(x,t) &= \mp c_s s(x,t) \\
	s(x,t) &= \frac{\sqrt{e}\zeta}{\rho_0\lambda} \left(x \pm c_s t\right) e^{-\frac{(x \pm c_s t)^2}{2 \lambda^2}} \label{eq:pw_s}
\end{align}
where $p$ is the \emph{total wave pressure}, $u$ the \emph{wave velocity}, $s$ the \emph{condensation},
$\zeta$ the \emph{wave amplitude} and $\lambda$ a steepness factor.
To examine the capability of the NSBC to absorb such a wave, a two-dimensional square domain of fluid is considered,
containing an initially inhomogeneous distribution of density and velocity, representing a plane wave.
To observe the behaviour of the NSBC under different angles of incidence,
the wave is configured to approach the boundary under an angle of $60 \degree$ with respect to the horizontal axis.
The domain consists of $128 \times 128$ lattice units and is fully bounded by NSBC's ($\gamma=0.6$ and $f_f=0.0$).
The wave properties are set to $\rho_0=1.0$, $\zeta=0.01$ and $\lambda=l/32$.
To approximate equation \eqref{eq:we_0}, the fluid viscosity is assumed to be zero (e.g. $\tau=0.5$).
The MRT relaxation rates have been chosen as $\omega_e=\omega_{\epsilon}=\omega_q=1.9$.
The pressure results of Figure~\ref{fig:iw_rs} show that the reflectivity of the NSBC is nearly zero.
\section{Test case 2 -- Convected vortex} \label{sub:tc_cv}
\begin{figure}[ht]
	\includegraphics[scale=1.0, page=2]{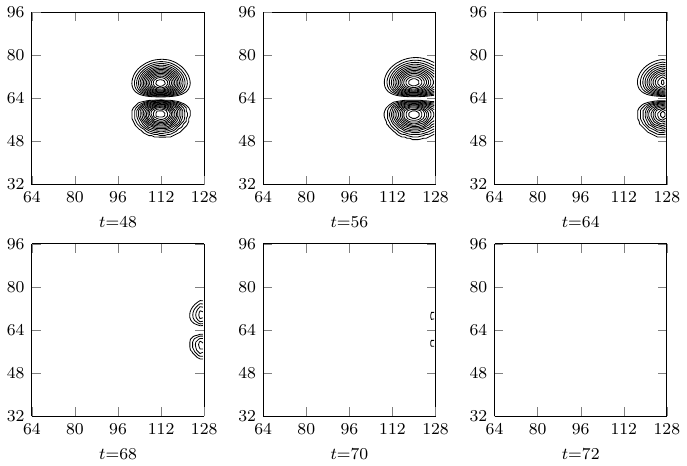}
	\caption{Test case 5: Convected vortex -- Isovalues of longitudinal velocity}
	\label{fig:cv_rs}
\end{figure}

Formed in stirred fluids, vortices are a major component in many flow applications.
Due to their characteristics and complex interaction with the surrounding fluid, absorption of vortices can be challenging.
To examine the capability of the NSBC in this area, a two-dimensional Lamb-Oseen vortex \cite{lamb1932} is convected towards the right boundary
of a square domain with a grid size of $128 \times 128$ lattice units \cite{wissocq2017}.
The vortex is initialized by introducing a local perturbation of the flow field according to:
\begin{align}
	u &= u_0 - \beta u_0 \frac{\left(y-y_0\right)}{R_c} e^{-\frac{r^2}{2 R_c}} \label{eq:cv_u} \\
	v &= \beta u_0 \frac{\left(x-x_0\right)}{R_c} e^{-\frac{r^2}{2 R_c}} \label{eq:cv_v} \\
	\rho &= \left[1-\frac{\left(\beta u_0\right)^2}{2 C_v} e^{-\frac{r^2}{2}}\right]^{\frac{1}{\gamma-1}} \label{eq:cv_d} \\
	r &= \left(x-x_0\right)^2+\left(y-y_0\right)^2 \label{eq:cv_r}
\end{align}
where $u_0=0.1$ is the reference velocity, $\beta=0.5$ a coefficient and $R_c=20$ the vortex radius (All quantities are in lattice units).
The \emph{gas constant} $\gamma$ and the \emph{volumetric heat capacity} $C_v$ are defined by:
\begin{equation}
	\gamma=\frac{d+2}{d} \qquad C_v=\frac{d}{2} c_s^2
\end{equation}
where $d$ is the number of spatial dimensions. The Reynolds number equals $\text{Re}=10^3$ and is based on $u_0$ and the size of the computational domain.
Concerning the domain boundaries; a Dirichlet velocity boundary \cite{lbm2017} is defined at the left and an NSBC with $\gamma=0.6$ and $f_f=0.0$ is defined at the right;
the bottom and top boundaries are assumed to be periodic. The MRT relaxation rates are $\omega_e=\omega_{\epsilon}=1.4$ and $\omega_q=1.2$.
The results from Figure~\ref{fig:cv_rs} show that the vortex is fully absorbed by the boundary.
%
%
%
%
%
\section{Summary} \label{sec:sum}
A new type of non-reflective boundary formulation (NSBC) is proposed, based on the approximately symmetrical shape of the non-equilibrium distribution function.
In this formulation, the incoming populations at a boundary node are assigned the non-equilibrium contributions of the outgoing populations.
The equilibrium contributions of these incoming populations are computed using a predicted macroscopic fluid vector, determined by a simple gradient based extrapolation method.
Additionally, a far-field flow criterion can be applied to this fluid vector, to allow for convergence towards a certain reference value.
After all populations have been assigned, they are rescaled and shifted with respect to the fluid vector, as to satisfy conservation of the macroscopic moments.
To examine the non-reflectiveness of proposed boundary condition, two different tests have been carried out.
In the first test the capability of the NSBC to absorb acoustic waves has been studied.
Results show that the reflections are nearly zero, even when considering a large angle of incidence.
As a second test the aborption of a convected vortex has been modelled.
Isovalues of the longitudinal velocity indicate that the vortex is completely absorbed by the boundary.
To summarize, the NSBC has found to be an interesting alternative for modelling non-reflective boundaries.
However further investigations are needed to determine the validity of present boundary formulation.

\begingroup
\hbadness 10000
\bibliographystyle{styles/spphys}
\bibliography{references}
\endgroup
\end{document}